\documentclass[aps,twocolumn,showpacs,preprintnumbers,amsmath,amssymb]{revtex4}

\usepackage{graphicx}
\usepackage{dcolumn}
\usepackage{bm}

\begin{document}

\preprint{submitted to PRL}

\title{Plasmonic bandgaps and Trapped Plasmons on Nanostructured Metal Surfaces}

\author{T.A. Kelf}
\email{tkelf@phys.soton.ac.uk}
\author{Y. Sugawara}%
\author{J.J. Baumberg}%
\affiliation{School of Physics and Astronomy, University of Southampton \\ Highfield, Southampton, SO17 1BJ}%

\author{M. Abdelsalam}%
\author{P.N. Bartlett}%
\affiliation{School of Chemistry, University of Southampton, Highfield, Southampton, SO17 1BJ}%

\date{\today}


\begin{abstract}
Nanostructured metal surfaces comprised of periodically arranged spherical voids are grown by electrochemical deposition through a self-assembled template.
Detailed measurements of the angle- and orientation-dependent reflectivity
reveal the spectral dispersion, from which we identify the presence
of both delocalized Bragg- and localized Mie-plasmons. These couple strongly
producing bonding and anti-bonding mixed plasmons with anomalous dispersion properties. Appropriate plasmon engineering of the void morphology selects the plasmon spatial and spectral positions, allowing these plasmonic crystal films to be optimised for a wide range of sensing applications.
\end{abstract}

\maketitle

In recent years the topic of surface plasmons and plasmon-active
surfaces has become of interest to a broad range of
disciplines.\cite{Barnes03}  Stemming from the desire to construct
plasmon devices and the ability to make and characterize metallic
structures on the nano-scale, plasmonics finds applications in
fields as diverse as optical switching\cite{Krasavin05}, near-field photonics\cite{Zayats03} and surface-enhanced Raman
spectroscopy.\cite{Tian02} A critical issue for the
application of plasmons is their location and spatial extent
within complicated 3D metal geometries. Despite impressive
progress in the field of scanning near-field microscopy, the technique is selective in the plasmons it records, is somewhat invasive to the plasmon fields, and
is unable to distinguish between localized or delocalized
modes.\cite{Hohng02} The difficulties are exemplified by the
extended debate about how light squeezes through arrays of
sub-wavelength holes\cite{Ebbesen98}.  Similarly, extreme localization
of plasmons on rough metal surfaces is thought to produce
`hot-spots' for resonant Raman scattering.\cite{Stockman01} It has, however, proved difficult to study in detail this process of plasmon localization.

Here we study the band-structure of, and coupling between
different plasmon modes on a series of carefully controlled
metal nanostructured surfaces. By measuring the polarized
reflectivity of a focused white light laser while varying the
sample orientation and incident angle, the spectral
dispersion of surface and localized plasmons as a function of
sample geometry is measured. Strong coupling between different
plasmon modes is clearly observed, along with the systematic
localization of surface plasmons at a critical geometrical
condition. Understanding the complexity of the plasmon
localization opens the way to plasmon engineered surfaces,
designed for particular applications.

Along the boundary between a metal and a dielectric,
surface plasmon-polaritons (SPPs) can propagate freely. By
patterning such surfaces with a periodic array of holes, an
incident optical field can be diffracted and couple to these SPPs.\cite{Hooper02}  The SPPs will also multiply scatter off the periodic components
of the array leading to `Bragg' plasmon modes with plasmonic
bandgaps. Plasmons can become localised if the field fluctuations
become spatially pinned and disconnected from each other. These
localised modes are highly dependent on the geometry of the
surface and have attracted a growing interest for achieving large
field enhancements.\cite{Bergman03}

The nano-structured surfaces investigated here are formed using a nano-casting process, by electrochemical
deposition through a template of self-assembled latex
spheres.\cite{Bartlett00,Bartlett03}  The resulting metallic mesh
reflects the order of the self-assembled close-packed template,
allowing convenient control of the pore diameters and regularity
of the array.  Templates are produced using a capillary force
method, allowing a monolayer of well-ordered spheres to be
produced, with sphere diameters, $d$, (and hence pitch)
between 100nm and $>10\mu$m. Electro-deposition while measuring the total charge passed allows the accurate growth of metal to a required thickness $t$.  Furthermore by systematically retracting the
sample from the plating bath during growth, the nano-structure
geometry can be graded. After deposition the template is
dissolved, leaving the freestanding structure. This allows the
production of shallow well-spaced dishes as well as encapsulated
spherical voids on a single sample. Optical and electron
microscopy shows that the resulting surfaces are smooth on the
sub-10nm scale. We combine these with scanning probe microscopy to
determine the film thickness locally.  Here we concentrate mostly on Au voids, although very
similar features are seen for Ag samples.

\begin{figure}
\includegraphics[width=85mm]{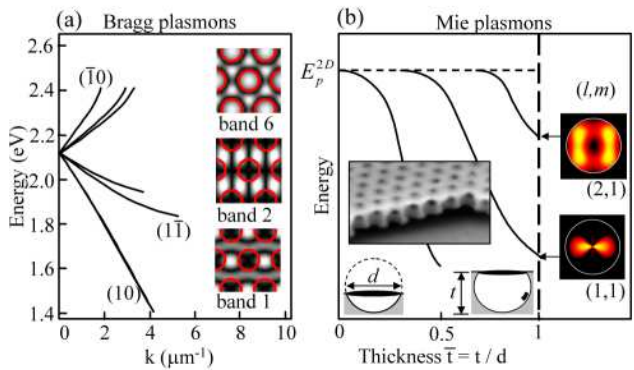}
\caption{\label{fig1} (a) Dispersion of zone-folded surface plasmon polaritons, for voids in Au spaced by 600nm. Images show surface electric field distribution for the corresponding standing wave modes at $k=1/\mu$m close to $\Gamma M$, red circles mark dishes. (b) Localised plasmon energies vs normalised sample thickness, $\bar{t}$ for a random array of separated voids. Calculated $xy$ field distributions for encapsulated spheres, $\bar{t}$=1. Inset: scanning electron micrograph of sample at $\bar{t}$=0.9.}
\end{figure}

On thin regions of the sample (normalised film thickness, $\bar{t}
= t/d < 0.2$) where the surface takes the form of a hexagonal array
of shallow dishes, SPPs are observed. These states multiply
scatter off the rims of the dishes, forming plasmonic bandgaps
similar to those formed in 2D dielectric photonic crystals.
For weak scattering these Bragg plasmons track the
folded plasmon dispersion in the 2D lattice (Fig.\ref{fig1}a).  The Bragg scattering mixes these six-fold degenerate modes, splitting their energies and forming travelling Bloch waves with different standing wave field distributions within each unit cell (bands 1,2,6 shown for small in-plane $k$ along the $\Gamma K$ direction)\cite{Joannopoulos95}. These distributions are calculated in a weak 2D dielectric photonic crystal approximation, and show markedly different overlap with the circular holes.
Because the Bragg plasmons are travelling Bloch modes, their energy depends strongly on their
direction of propagation.\cite{Satpathy90} 
As the thickness of the film increases,
the rim of the spherical voids expands while the pitch remains
constant, strongly increasing the scattering of the Bragg
plasmons. For hemispherical voids, the top planar surface breaks
up into disconnected triangular islands, preventing
straightforward passage of the surface modes.

On the other hand, localized plasmons are trapped in the deep
spherical cavities embedded in the film at larger thicknesses.
Within fully encapsulated voids these modes can be modelled using
a Mie scattering approach \cite{Coyle01,Teperik05},
hereafter referred to as `Mie' plasmons. Because they are
localized they can directly couple to optical fields and are
nearly isotropic - largely uninfluenced by the electric fields from
neighbouring voids. When the voids are truncated ($\bar{t}<1$),
Mie modes  of different symmetry are coupled together and their energies are forced upward (Fig.\ref{fig1}b) towards the energy of the 2D surface plasmon (the
flat metal limit for $t$=0).  The lines are extracted from experiments on amorphous random arrays of separated voids 
of the same dimensions (as previously identified in \cite{Coyle01}). 
The Mie plasmon wavefunctions correspond
to spherical harmonics similar to atomic orbitals, labelled by angular and azimuthal quantum
numbers, ($l,m$).\cite{Coyle01} For a single 600nm spherical void in Au, the
energies of $p$ and $d$ Mie states are 1.64eV and 2.22eV
respectively, compared to the Bragg plasmon mode at $k$=0 of
2.12eV.

Hence as the thickness of the film is increased, the localized Mie
modes drop from the 2D plasmon around 2.5eV, down through the
propagating Bragg modes. It is the strong coupling between these
modes that controls the plasmon properties on these nanostructured
metal surfaces.

The key role of spectral- and angular-dispersion motivates the
collection of accurate data from different thicknesses of these nanostructures. An
automated goniometer configuration allows precise control over the
orientation and position of a sample, which is illuminated by a super-continuum white light laser with wavelengths from 480nm up to
2$\mu$m. The reflected light is collimated then collected using a 100$\mu$m
diameter multimode fibre and analysed in a broadband
spectrometer.  The use of a high-brightness laser allows high angular resolution
from a small sample region.\cite{Netti00}  Using suitable polarization optics, both polarization-preserving and polarization-rotating spectra are observed.  These show similar trends in absorption for both TE and TM incident light due to the strong polarisation conversion from the voids.\cite{Coyle03} Here we concentrate on plasmon absorption dips in
the co-polarized spectra, leaving the more subtle
polarization-rotation spectra to a subsequent publication.

\begin{figure}
\includegraphics[width=85mm]{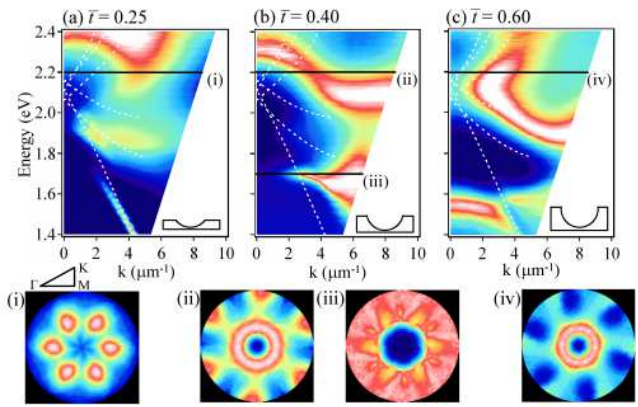}
\caption{\label{fig2}Measured energy dispersion of reflectivity for TM polarised light as a function of in-plane wavevector for increasing void thickness, $\bar{t}$ (a-c). Log colour scale, white dotted lines show zone-folded plasmon dispersion, sample orientations of $\phi$ = $30^{\circ}$ in all cases. (i-iv) $k$-space cuts through dispersion relation at; (i) ($\bar{t}$,E)=(0.25,2.2eV); (ii) ($\bar{t}$,E)=(0.4,2.2eV); (iii) ($\bar{t}$,E)=(0.4,1.7eV); (iV) ($\bar{t}$,E)=(0.6,2.2eV), symmetery shown above (i).  Light shade corresponds to absorption features.}
\end{figure}

To exemplify the physics of the plasmons, we focus on a gold
nano-void film of pitch 600nm, which most clearly resolves the plasmon
interactions in the visible spectral region. For each nanostructure morphology
on this graded-thickness sample, the spectral
absorption is measured over a range of incident angles and sample
orientations. Figure \ref{fig2} shows cuts through these data sets of reflectivity,
$R(\omega,k_x,k_y)$ for three different film thicknesses. The
upper images show the energy dispersion with incident angle, measured in the azimuthal $\Gamma$ M direction.
The dispersion and coupling strength of various plasmon modes is clearly observed, along with anticrossings between different modes.  The lower images are slices at fixed energy of the
symmetry of the modes in the $k_x-k_y$ plane. Propagating modes
will see different dispersions in different directions and hence
posses a 6-fold symmetry arising from the hexagonal lattice. Localized
plasmon modes are confined within individual voids, so feel little
of the samples asymmetries, appearing almost circular in these plots.

On thin nano-void films (Fig.\ref{fig2}a) nearly-pure Bragg modes are
observed. This shows that for shallow dishes, the observed modes
are indeed propagating plasmons, which, even at high energies
(where the imaginary part of the Au dielectric constant increases
strongly near the 2D plasmon energy), propagate for many lattice
periods through the structure. Although they conform well to the
folded Bragg dispersion theory (dotted), 
plasmon mode repulsion is clearly identified in the lower modes near normal incidence,
corresponding to a plasmonic energy bandgap at $k$=0. This bandgap is
reduced for even thinner films, but shows already that plasmon
scattering is extremely strong.

As the voids become deeper the Mie plasmon drops in energy and approaches the Bragg plasmon,
producing a mixing of the states (Fig.\ref{fig2}b). The new mixed modes become highly non-dispersive, and
their mode-symmetries appear more localized. In particular the lower
Bragg modes are mixed with a Mie mode and form new states separated
by an energy gap of $>$400meV. This strong-coupling regime implies
that the Bragg modes are strongly perturbed and mixed when the Mie
states within the voids are in resonance with them.  The modes
visible in Fig.\ref{fig2}b smoothly convert from Mie to Bragg in the top
band and vice versa in the lower band.  As the thickness increases
further the next Mie mode appears at higher energy, while the lowest Mie mode drops below the Bragg energy and the surface modes begin to reform (Fig.\ref{fig2}c). As the modes come out of resonance, the weaker coupling gives small anticrossings between lower Mie plasmon in the void and the lower Bragg plasmon which moves on the flat surface (at 1.5eV in Fig.\ref{fig2}c). The coupling of the mixed plasmon mode with incoming light is modified in the strong Bragg-Mie coupling regime. The reflectivity measures the plasmon dispersion modulated by this coupling integral. Out of resonance,
the Mie modes are visible at normal incidence (e.g. for
$\bar{t}=0.25$) as are the Bragg modes for smaller $\bar{t}<0.15$
(not shown). However when mixed, some of the resulting modes can be
conspicuously absent around $k<1 \mu$m$^{-1}$ (Fig.\ref{fig2}c). This implies that
either the wavefunction becomes confined into a ring in $k$-space,
or that the upper mixed plasmon is not dipolar in this symmetry. The
former explanation is unlikely, given the circular dispersion
symmetry observed since the plasmon would need to be
delocalized over more than twenty spherical voids. The latter
explanation corresponds to an anti-symmetric combination of
the two dipole-allowed Bragg and Mie modes, which gives rise to
a `dark' quadrupolar plasmon state at $k$=0.

\begin{figure}
\includegraphics[width=85mm]{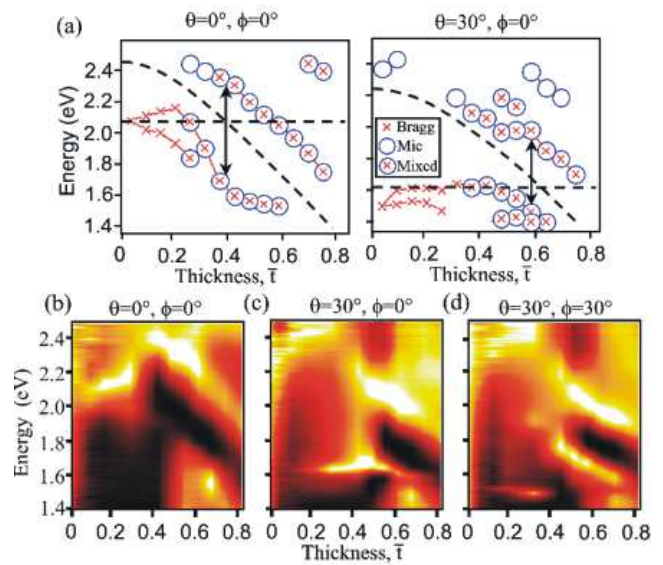}
\caption{\label{fig3} (a) Extracted energies of Bragg and Mie plasmons as a function of normalised sample thickness for $\theta=0^{\circ},30^{\circ}$, showing anticrossing (dashed and solid lines show experimentally determined pure Bragg (for $\bar{t}$ = 0) and Mie mode behaviour respectively). (b-d) Reflection image plots of energy vs normalised thickness for, (b) $\theta=0^{\circ},\phi=0^{\circ}$, (c) $\theta=30^{\circ},\phi=0^{\circ}$, (d) $\theta=30^{\circ},\phi=30^{\circ}$.}
\end{figure}

The resonant coupling of the Bragg modes to a given Mie mode is
dependent on their in-plane momentum. Extracting the mode energies
at $\theta=0,30^{\circ}$ (Fig.\ref{fig3}a) shows that the maximum Rabi
splittings of the mixed modes occur at different thicknesses.  The nature of a given node is assertained by observing its $\theta$,  $\phi$ dispersion. At normal incidence the Bragg plasmons are near degenerate close to
the Bragg energy at 2.1eV.

At increasing thickness, the Mie mode comes into resonance and an
anticrossing is observed at $\bar{t}=0.4$. When off
normal incidence, the six Bragg modes open up, of which the lower
two are clearly visible. One of these modes is quickly suppressed
by $\bar{t}=0.3$, while the other shows a Rabi splitting
maximized at $\bar{t}>0.6$. Thus the downwards angular
dispersion of both Mie and Bragg modes causes the maximum
coupling to occur for different $\bar{t}$ at different angles. The
observed splitting of the plasmons corresponds to
bonding and antibonding plasmon orbitals. We note however 
that the coupling is between a {\it localised} and a {\it
delocalised} plasmon, and thus rather different to electronic states in
molecules.

The gap between bonding and anti-bonding mixed plasmons can be
seen more clearly in (Fig.\ref{fig3}b-d).  The three images show the energy dependence of the different modes at increasing void thickness, tracked through the reflectivity, for incident angles of 0$^{\circ}$,30$^{\circ}$.  In the latter case we show the thickness dependence for both principal sample orientations. In all cases the drop in energy of the Mie modes can be
clearly observed with increasing thickness. At normal incidence, (Fig.\ref{fig3}b) the Bragg modes visible around 2.1eV abruptly disappear at $\bar{t}=0.4$ while a localized Mie state appears at 1.55eV for $\bar{t}>0.5$. For non-normal incidence, the mode mixing depends on $\phi$. Examining the two orientations for
plasmons travelling either along the necks between voids 
($\phi=0^{\circ}$, Fig.\ref{fig3}c) or along the line of voids
($\phi=30^{\circ}$, Fig.\ref{fig3}d) reveals the symmetry of the Bragg modes that mix with the plasmons trapped in the voids. In both directions, the lowest Bragg mode (band 1) abruptly vanishes
for $\bar{t}>0.4$. For $\phi=0^{\circ}$, the next Bragg mode (band 2) around 1.6eV
strongly couples to the Mie mode forming a distinctive band gap. Along
$\phi=30^{\circ}$ a high energy mode emerges and drops down
into the lowest localized mode at 1.55eV, producing the same band gap. Examining the field
distributions suggest that this is the upper Bragg mode (band 6) which has its largest electric field already within the void (Fig.\ref{fig1}a).

\begin{figure}
\includegraphics[width=85mm]{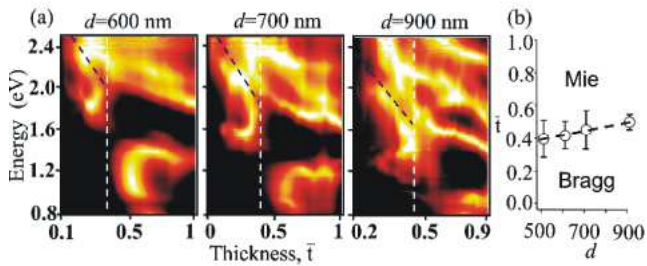}
\caption{\label{fig4} (a) Energy dispersion vs thickness of Au nanovoid samples of different diameter $d$, mapped by reflectivity of TM polarised light for sample orientations $\theta=30^{\circ}$, $\phi=0^{\circ}$. Light shade correspond to absorption features.  Vertical white line shows position of Bragg mode collapse, blue line indicates separation between Bragg modes (low energy) and Mie modes (high energy). (b) Extracted sample thickness (normalised) of Bragg mode collapse with increasing void diameter.}
\end{figure}

Thus the scenario of localization on these nanostructured surfaces
can be summarized: Bragg plasmons existing in the plasmonic
crystal of shallow dishes couple strongly to Mie modes forming new
mixed states. Particular Bragg modes whose fields are already
stronger in the dishes lose their extended character and drop into
Mie plasmons. Bragg modes with fields on the surface vanish at a
critical thickness.

These conclusions are confirmed by experiments
on a large number of samples of different pitch (and void
diameter), and composition. In each case, the Bragg plasmons vanish at a critical
thickness just as the lowest localized modes appear (Fig.\ref{fig4}).
Currently, exact calculation of plasmon energies in
arrays of truncated spherical voids is not yet possible, though
progress is being made.\cite{Teperik05} However, it is interesting to
see already the universal features of the experimental plasmon
phase diagram (Fig.\ref{fig4}b). Plasmon transport is rather different in
the regimes of delocalized Bragg modes (coherent plasmon transport
with scattering and damping) and localized Mie modes (hopping
plasmon transport, enabled by coupling to nearby Bragg modes).

As the Mie modes localize the plasmons, the surface electric fields
increase in strength, thus enabling many applications for plasmonic
devices. For instance, we have recently shown that these surfaces
are ideal for surface enhanced Raman scattering (SERS) which is
strongly enhanced in the localised Mie regime.\cite{Abdelsalam05}
The ability to now predict and engineer appropriate plasmons
suited to specific pump and emission wavelengths is a drastic
improvement on previous SERS substrates. This approach has proved
so promising that it is now in commercial production for a variety
of applications in medical and pharmaceutical screening, homeland
security and environmental monitoring.\cite{Mesophotonics}
Understanding plasmons on nanostructured surfaces is the gateway
to a new generation of nano-plasmonic devices.

We enthusiastically acknowledge discussions with Tatiana Teperick,
Sacha Popov, and Javier Garcia de Abajo. This work was supported
by EPSRC grants EP/C511786/1, GR/R54194/01. YS was supported by a 
{\it JSPS} Fellowship for Research Abroad, HEISEI 15.

\end{document}